\documentclass[traditabstract]{aa}
\usepackage{amssymb,amsmath,natbib,graphicx,rotating}
\usepackage[format=hang]{caption,subfig}
\bibpunct{(}{)}{,}{a}{}{,}

\title{\bf A study of the color diversity around maximum light in Type Ia supernovae}
\author{R\'egis Cartier \inst{\ref{affil1}, \ref{affil2}} \and Francisco F\"orster \inst{\ref{affil1}} \and Paolo Coppi \inst{\ref{affil2}} \and Mario Hamuy \inst{\ref{affil1}} \and Keiichi Maeda \inst{\ref{affil3}} \and Giuliano Pignata \inst{\ref{affil4}} \and Gaston Folatelli \inst{\ref{affil3}} }
\institute{Departamento de Astronom\'\i a, Universidad de Chile, Casilla 36-D, Santiago, Chile \label{affil1}  
\and Department of Astronomy, Yale University, New Haven, CT 06520, USA \label{affil2}
\and Institute for the Physics and Mathematics of the Universe (IPMU), University of Tokyo, 5-1-5 Kashiwanoha, Kashiwa, Chiba 277-8583, Japan \label{affil3}
\and Departamento Ciencias F\'isicas, Universidad Andres Bello, Av. Rep\'ublica 252, Santiago, Chile \label{affil4}
}

\abstract{

\noindent

From a sample of 12 well-observed Type Ia supernovae, we find clear 
evidence of correlations between early phase $(U-B)$, $(V-R)$, and $(V-I)$
colors and the velocity shifts of the $[{\rm Fe~II}]$ $\lambda 7155$
and $[{\rm Ni~II}]$ $\lambda 7378$ nebular lines measured from late-phase spectra.
As these lines are thought to trace the ashes of the initial deflagration
process, our findings provide additional support to the new
paradigm of off-center explosions in Type Ia supernovae, and we interpret these
correlations as viewing angle effects in the observed colors. We also show that
the nebular velocity shifts are related to the strength and
width of the Ca II $H\&K$ and IR-triplet lines near-maximum light.
The evidence presented here implies that the viewing angle must be taken into
account when deriving extinction values and distances in future cosmological
studies based on Type Ia supernovae.

}


\begin{document}

\authorrunning{Cartier et al.}
\titlerunning{A Study Of The Color Diversity Around Maximum Light In Type Ia Supernovae.}

\maketitle

\section{INTRODUCTION}

Type Ia Supernovae (SNe) play an important role in modern astrophysics thanks to their
great power in extragalactic distance determinations, which permitted the discovery of the cosmic
acceleration \citep{riess98, perl99}. Line-of-sight extinction has to be taken into account to 
correct the distance estimate of the SN host galaxy. Several empirical methods that
make use of SN colors have been developed for this purpose and very different extinction laws from the Galactic
law need to be invoked to reduce the dispersion in the calibration of SNe Ia in
cosmological studies \citep{hicken09, folatelli10, wang09b}.

Theoretical and observational studies have shown evidence of an
off-center deflagration  as the
initial burning process in SNe Ia \citep{kasen09, maeda10a, maeda10c, maeda11}.
\citet{maeda10c} show a
relation between the velocity gradient of the Si II lines observed near maximum light
and the nebular velocity shifts
measured from the $[{\rm Fe~II}]$ $\lambda 7155$ and $[{\rm Ni~II}]$ $\lambda 7378$ lines,
which are thought to trace the region where the initial deflagration occurs.
The off-center explosion model of \citet{maeda10c} explains the large diversity in velocity gradients near
maximum light described in \citet{benetti05}, who classifies as either high velocity gradient (HVG) or low velocity
gradient (LVG) SNe depending on the velocity gradient of the $Si~II~\lambda 6355$ line.

\citet{kasen09}, \citet{maeda11}, and \citet{foley11} demonstrate both theoretically and observationally 
that the $(B-V)$ colors and luminosity depend on the asymmetry of the
explosion and the viewing angle. Here we present additional evidence that favors
this new paradigm of asymmetric explosions in the form of strong correlations of $(V-I)$, $(V-R)$, and $(U-B)$ colors 
at very early phases with the nebular velocity shifts at late epochs. These correlations
could be useful in improving SN color calibrations in future cosmological studies.

\section{SAMPLE AND METHODOLOGY}
\label{meth}

Our sample consists of 12 nearby SNe with a well-sampled light curve,
photometry from at least six days before maximum light and late-time
spectroscopic observations, i.e. at least 150 days after maximum, to 
measure nebular velocity shifts. In Table \ref{Sample_tab}, we
summarize the decline rate ($\Delta m^{B}_{15}$)
and line-of-sight reddening values of our Type Ia SNe sample.

\begin{table}
\caption{Sample of Type Ia Supernovae.\label{Sample_tab}}
\centering
\begin{tabular}{ccccc}
\hline
SN & $\Delta m^{B}_{15}$ & \emph{E(B-V)}$_{Gal}$ & \emph{E(B-V)}$_{host}$ & Ref. \\
\hline
1990N  & $1.04\pm 0.14$ & $0.026$ & $0.163 \pm 0.066$  & L98  \\
1994D  & $1.40\pm 0.15$ & $0.022$ & $-0.028 \pm 0.046$ & P96 \\
1998aq & $1.06\pm 0.12$ & $0.014$ & $0.060 \pm 0.025$  & R05 \\
1998bu & $1.04\pm 0.19$ & $0.025$ & $0.401 \pm 0.017$  & J99 \\
2001el & $1.16\pm 0.11$ & $0.014$ & $0.318 \pm 0.020$  & K03 \\
2002bo & $1.02\pm 0.23$ & $0.025$ & $0.437 \pm 0.051$  & B04 \\
2002dj & $1.14\pm 0.17$ & $0.096$ & $0.149 \pm 0.050$  & P08 \\
2002er & $1.25\pm 0.15$ & $0.157$ & $0.189 \pm 0.057$  & P04 \\
2003du & $1.00\pm 0.11$ & $0.010$ & $0.032 \pm 0.020$  & S07 \\
2004eo & $1.31\pm 0.16$ & $0.109$ & $0.048 \pm 0.024$  & P07 \\
2005cf & $1.00\pm 0.13$ & $0.097$ & $0.138 \pm 0.012$  & W09 \\
2006dd &$1.12\pm 0.18$ & $0.021$  & $0.054 \pm 0.019$  & S10 \\
\hline
\end{tabular}
\tablebib{B04. \citet{benetti04}; J99. \citet{jha99}; K03. \citet{kris03}; L98. \citet{lira98};
  P96. \citet{patat96}; P04. \citet{pignata04}; P07. \citet{pastorello07};
  P08. \citet{pignata08}; R05. \citet{riess05}; S07. \citet{stanishev07}; S10. \citet{stritzinger10};
  W09. \citet{wang09a}}
\end{table}

\begin{figure}
\centering
\includegraphics[width=0.4 \textwidth, height=0.4 \textwidth]{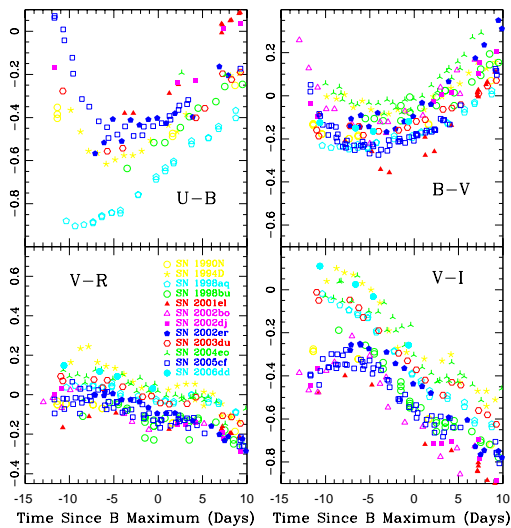}
\caption{Reddening-corrected colors of our sample of Type Ia supernovae.}
\label{ColorsMax_fig}
\end{figure}

We estimated the epoch of maximum light for each filter, $\Delta
m^{B}_{15}$, and the SN colors at different epochs using fifth order
polynomial interpolated light curves.  We used K-corrected photometry
for SN~2004eo and S-corrected photometry whenever available in
the literature. Given that we are interested in assessing the
relations between optical colors and nebular velocity shifts around
maximum light, we estimated host galaxy reddening using the Lira
relation \citep{phillips99}, which is a method independent of SN
colors at maximum. To estimate Galactic extinction, we used the maps of
\citet{schlegel98} and the extinction law ($R_V = 3.1$) of
\citet{cardelli89}, which is also used for host galaxy reddening. The
reddening estimates are shown in Table \ref{Sample_tab} and the
corresponding reddening-corrected light curves are shown in
Figure~\ref{ColorsMax_fig}. We adopted the nebular velocity shifts
($V_{\rm neb}$) from \citet{maeda11} and studied their relation with SN
colors at different epochs.

Finally, we investigate the effect of Ca II lines, which are the
dominant features in the $U$ and $I$ bands, in the colors of SNe Ia.
To quantify the effects of Ca II in the SN colors, we model the
available spectra acquired four days priot to maximum using the code syn++
described in \citet{thomas10}, and then we computed synthetic
photometry for our models with and without Ca II lines.

\section{OPTICAL COLORS AND THE NEBULAR VELOCITY SHIFTS}
\label{colvsvneb}

We now study the relation between optical colors and viewing angle in
the context of off--center ignition models \citep{kasen09, maeda10a,
  maeda10b, maeda10c}. We assume that the viewing angle is traced by
the nebular velocity shift of the $[{\rm Fe~II}]$ $\lambda 7155$ and
$[{\rm Ni~II}]$ $\lambda 7378$ lines, which are suggested to originate
from the ashes of the deflagration process \citep{maeda10a, maeda10c,
  maeda11}.

We first note that in Figure~\ref{ColorsMax_fig} the $(U-B)$ and
$(V-I)$ colors show a large diversity before and near maximum
light. The $(B-V)$ and $(V-R)$ color curves, on the other hand, are
more homogeneous at all epochs.  We show that some of this
diversity in colors could be due to the viewing angle.

We now introduce zero subscripts for the reddening-corrected colors. To 
quantify possible correlations between unreddened colors and $V_{\rm neb}$, 
we use the linear model:

\begin{equation}
(color)_0 = \alpha + \beta \times (V_{\rm neb} ~/~ 1000 {\rm ~km
    ~s^{-1}}).
\label{V_I_Vneb}
\end{equation}

\begin{figure}
\centering
\includegraphics[width=0.4 \textwidth, height=0.4 \textwidth]{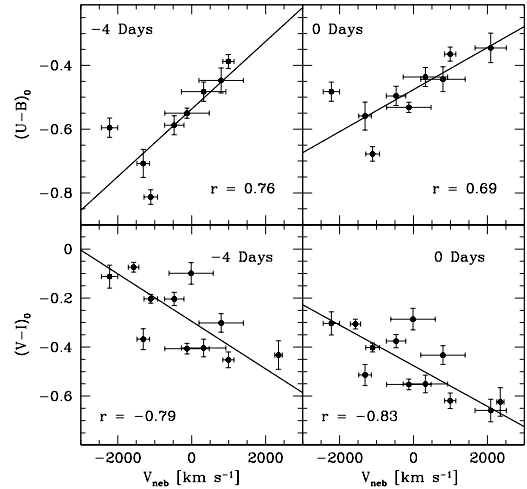}
\caption{(top) $(U-B)_0$ vs $V_{\rm neb}$ at two different epochs. $r$ is the linear correlation coeficient.
(bottom) same as above but for $(V-I)_0$.
}
\label{Vnebvscolor_fig}
\end{figure}

In Figure~\ref{Vnebvscolor_fig}, we present the evolution of the
$(U-B)_0$ and $(V-I)_0$ colors as a function of $V_{\rm neb}$, at -4
and 0 days since maximum light, as well as the best--fitting linear
model from eq.~(\ref{V_I_Vneb}). From the top and bottom panels, it is
clear that there is a linear correlation between $(U-B)_0$ and $V_{\rm
  neb}$ and a linear anticorrelation between $(V-I)_0$ and $V_{\rm
  neb}$. Although not shown here, these correlations are clearly
present from -8 to +8 days since maximum light, going generally from steeper to
shallower slopes with time, in agreement with the scatter seen in 
Figure~\ref{ColorsMax_fig}. A weaker, but significant
anticorrelation was found between $(V-R)_0$ and $V_{\rm neb}$, and no
significant correlation was found between $(B-V)_0$ and $V_{\rm
  neb}$. Best--fitting parameters for eq.~(\ref{V_I_Vneb}) at selected
colors and epochs are shown in Table~\ref{model}. All the correlations
found are significant, with a very low probability of the data being
drawn from an uncorrelated population in all cases.

\begin{table}
\caption{Correlations between unreddened colors and $V_{\rm
    neb}$\label{model}} \centering
\begin{tabular}{ccccccc}
\hline
color & epoch & $\alpha$ & $\beta$ & $r$ \\
\hline
$(U-B)_0$ & -4 & $-0.537 \pm 0.015$ & $0.106 \pm 0.002$  & 0.76 \\ 
$(U-B)_0$ & 0  & $-0.476 \pm 0.012$ & $0.066 \pm 0.011$  & 0.69 \\ 
$(U-B)_0$ & +4 & $-0.320 \pm 0.013$ & $0.079 \pm 0.015$  & 0.71 \\
$(V-I)_0$ & -4 & $-0.295 \pm 0.014$ & $-0.097 \pm 0.015$ & -0.79 \\ 
$(V-I)_0$ & 0  & $-0.476 \pm 0.012$ & $-0.083 \pm 0.012$ & -0.83 \\ 
$(V-I)_0$ & +4 & $-0.596 \pm 0.013$ & $-0.076 \pm 0.014$ & -0.76 \\       
$(V-R)_0$ & 0  & $-0.071 \pm 0.010$ & $-0.038 \pm 0.008$ & -0.78 \\
\hline
\end{tabular}
\tablebib{Fitting parameters for the model described with
  eq.~(\ref{V_I_Vneb}) for selected colors and epochs. The last column
  contains the Pearson correlation coefficient, whose values imply low
  probabilities of the data being drawn from an uncorrelated
  population: 0.029, 0.04, 0.032, 0.004, 0.001, 0.007, and 0.003 for each row,
  respectively.}
\end{table}

\section{EFFECT OF VIEWING ANGLE IN Ca II LINES}

Given that the Ca II $H\&K$ and IR-triplet are the strongest lines in
the $U$ and $I$ filters, respectively, it is pertinent to investigate
whether or not the strength of these features could be responsible for
the $(U-B)_0$ and $(V-I)_0$ variations and their dependences on $V_{\rm neb}$,
presented above. For this purpose, we use the $[{\rm Fe~II}]~\lambda 7155$ and
$[{\rm Ni~II}]~\lambda 7378$ line velocity shifts
as a proxy for viewing angle.

\begin{figure}
\centering
\includegraphics[width=0.4 \textwidth, height=0.4 \textwidth]{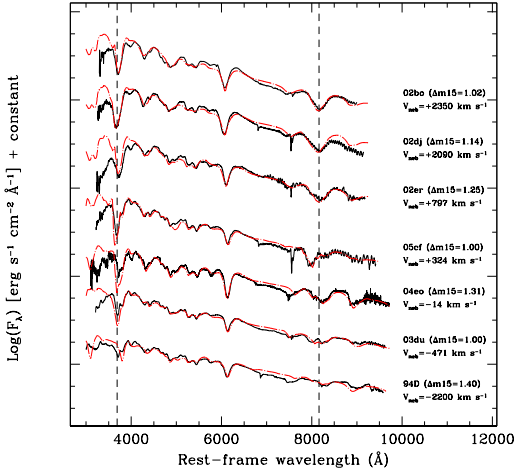}
\caption{Spectral evolution at -4 days since maximum as a function of nebular velocity shift ($V_{\rm neb}$).
The observed spectra in black and syn++ models in red (dot-dashed lines). The spectra are redshift 
and extinction corrected ($R_V = 3.1$). The dashed lines correspond to the expected rest-frame
wavelength of $H\&K$ and IR-triplet Ca II lines and the spectra in the figure correspond to
SN~1994D \citep{patat96}, SN~2002bo \citep{benetti04}, SN~2002dj \citep{pignata08}, SN~2002er \citep{kotak05}, SN~2003du \citep{sollerman04}, SN~2004eo \citep{pastorello07}, SN~2005cf \citep{bufano09}.}
\label{VnebvsCaII_fig}
\end{figure}

Figure~\ref{VnebvsCaII_fig} shows the optical spectra of our sample of
Type Ia SNe sbout four days before maximum, organized as a function of
$V_{\rm neb}$. The spectra in Figure~\ref{VnebvsCaII_fig} are
corrected for extinction using the \citet{cardelli89} Galactic
extinction law ($R_V = 3.1$) and the reddening estimates given in
Table \ref{Sample_tab}, shifted to $z=0$.

A clear trend in the width and strength of the Ca II $H\&K$ and
IR-triplet lines with nebular velocity shift is evident in Figure~\ref{VnebvsCaII_fig}.
For negative values of the nebular velocity shift, (i.e. seen from the side closer to 
the deflagration center in the \citealt{maeda10c} model), the Ca II lines are weaker, whereas these
features become stronger with increasing values for the nebular velocity
shift, (i.e. seen from the opposite side of the initial deflagration
in the \citealt{maeda10c} model).

At first glance, it seems that the strength of the Ca II $H\&K$ and IR-triplet
could explain, at least in part, the color variations with $V_{\rm neb}$. For example,
as $V_{\rm neb}$ increases and the Ca II $H\&K$ line gets stronger, the
$U$-band flux gets relatively weaker, making the SN look redder in $(U-B)_0$, in qualitative
agreement with the trend seen in the top-left panel of Figure~\ref{Vnebvscolor_fig}.
Likewise, as $V_{\rm neb}$ increases and the Ca II triplet gets stronger, the
$I$-band flux gets relatively weaker, making the SN look bluer in $(V-I)_0$, in qualitative
agreement with the trend seen in the bottom-left panel of Figure~\ref{Vnebvscolor_fig}.

To assess quantitatively the effect of Ca II lines in the colors of SNe Ia, we model
the spectra from Figure~\ref{VnebvsCaII_fig} using syn++ \citep{thomas10}.
In syn++, one computes a spectrum by specifying the location and
optical depth for a given set of ions. The input parameters for syn++
are the photospheric velocity, the optical depth, the
e-folding length of the opacity profile, the maximum and minimum
cut-off velocity for each ion, and the Boltzmann excitation temperature
for parameterizing line strengths. The spectral models are very good over all 
wavelengths except blueward of 3800 \AA, where the spectrum is dominated by a 
superposition of many metal lines.

\begin{figure}
\centering
\includegraphics[width=0.25 \textwidth, height=0.25 \textwidth]{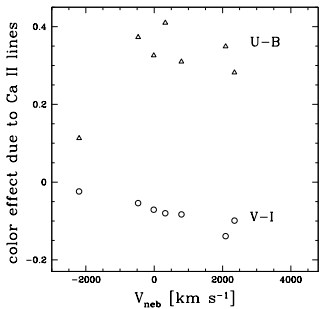}
\caption{$(U-B)_0$ and $(V-I)_0$ color differences from models of the spectra shown in
Figure~\ref{VnebvsCaII_fig}, and the same spectra after removing the $H\&K$ and IR-triplet
Ca II lines.
}
\label{colors_synphot_fig}
\end{figure}

From the resulting model spectra, we compute synthetic colors using
the $UBVRI$ filter functions given by \citet{bessell90}. We then
repeat the calculation after removing the Ca lines from the model
spectra.  Figure~\ref{colors_synphot_fig} shows the total effect of
the Ca II lines in the $(U-B)_0$ and $(V-I)_0$ colors, as a function
of $V_{\rm neb}$. We note that this is the ${\it
  maximum}$ possible effect because we then suppress all of the lines 
in the spectra. In $(U-B)_0$, the synthetic colors vary by $\sim$0.3 
mag, which could account for part of the observed variation of $\sim$0.4 
mag shown in Figure~\ref{Vnebvscolor_fig}. We do not see the
regular trend with $V_{neb}$ expected if Ca lines alone were responsible
for the correlations found in this work (eq.~\ref{V_I_Vneb}). However given the
difficulties in modeling the spectra shortward of 3800 $\AA$ it is
difficult to separate the contribution of Ca II lines in the colors.
In $(V-I)_0$, we recover the trend seen in
Figure~\ref{Vnebvscolor_fig}, although the overall variation in the
synthetic colors is a factor of four smaller than the observed
variation of $\sim$0.4 mag.  We conclude that the strength of the Ca
II lines is only partially responsible for the observed variations in
colors with $V_{\rm neb}$, and part of the color diversity is likely
due to differences in the slope of the continuum.

As expected, our synthetic $(B-V)_0$ and $(V-R)_0$ colors are weakly
affected by the Ca II lines, which fall at the edge of the B and R bands. 
In $(B-V)_0$, the color becomes slightly ($\sim$0.05) bluer as $V_{\rm neb}$ 
increases from -2000 to 2000  ${\rm ~km ~s^{-1}}$, which is caused by 
the emission component of the Ca II $H\&K$ P-Cygni profile. In $(V-R)_0$, 
the effect amounts to only 0.01 mag over the entire range of $V_{\rm neb}$.

From our syn++ models, we also find that to properly model the Ca II
IR-triplet of the LVG SNe in Figure~\ref{VnebvsCaII_fig} it is
necessary to add a detached high velocity Ca II component. In the case
of the HVG SNe (i.e. SN~2002er, SN~2002bo, and SN~2002dj), using high
optical depths and e-folding lengths with a single Ca II component can
result in good line models. As noted in \citet{foley11}, as the ejecta
velocity increases, the width in velocity space of the line-forming
region increases and the lines become broader. Therefore, in HVG SNe
individual lines overlap and can be modeled with a single broad Ca II
component.

\section{DISCUSSION AND CONCLUSIONS}

The theoretical work of \citet{kasen09} has put forward the notion
of intrinsically asymmetric ignition and subsequent detonation in Type
Ia SNe, to explain the ``peak luminosity-decline rate
relation" and color diversities for a given decline rate.  Strong
support for this idea was provided by \citet{maeda10c} by
successfully relating the early Type Ia velocity gradient diversity
described in \citet{benetti05} with the late-time nebular velocity
shifts of the $[{\rm Fe~II}]$ $\lambda 7155$ and $[{\rm Ni~II}]$
$\lambda 7378$ lines, interpreting the nebular velocity shifts as an
indication of viewing angle for an initially off-center deflagration.

We have presented clear evidence of correlations between the
late-time nebular velocity shifts, or viewing angle, and the
$(U-B)_0$, $(V-R)_0$ and $(V-I)_0$ reddening--corrected colors from -8
to +8 days since B-band maximum, providing additional support to the
new paradigm of off-center explosions. In $(B-V)_0$, on the other
hand, we find no significant correlation with the nebular velocity
shifts. The strengths of these color dependences on $V_{\rm
  neb}$ imply that the ejecta are asymmetric and that they are as 
important as the relations between color and $\Delta m^{B}_{15}$.

We also note that the $(B_{max} - V_{max})$ color dependence on velocity gradient
found in \citet{foley11} from a sample of 121 SNe Ia \citep{wang09b} could not be investigated
in our sample because of the small number of SNe. On the other hand, the
correlation between extinction and $\Delta m^{B}_{15}$ corrected $(B-V)$
colors and $V_{\rm neb}$ found in \citet{maeda11} using the relations
by \citet{folatelli10} are recovered in our sample. Our estimate of the host galaxy 
extinction is systematically higher by 0.08 mag than the values given in \citet{maeda11}. 
These differences are possibly caused by our use of the Lira relation, which is based on the 
late-time $(B-V)$ colors, while they employ the $(B_{max} - V_{max})$ pseudo color at
maximum \citep{folatelli10}. We do not find any relation between the residuals of our 
$(B_{max} - V_{max})$ colors or extinction estimates and the \citet{maeda11}
estimates with $\Delta m^{B}_{15}$ or $V_{\rm neb}$.

The reader may wonder whether our results are an artifact of our method 
for inferring extinction corrections. This concern can be ruled out
given that the $(U-B)_0$ vs $V_{\rm neb}$ and $(V-I)_0$ vs $V_{\rm neb}$ correlations
go in opposite directions, so that it is impossible to wash out both 
relations simultaneously by changing the reddening technique.

We also show that there is a relation between $V_{\rm neb}$
and the shape and strength of the Ca II $H\&K$ and IR triplet lines,
which are the strongest features in the $U$ and $I$ bands,
independently of $\Delta m^{B}_{15}$. This agrees with
\citet{tanaka08}, who noted that HVG SNe tend to have stronger Ca II
features. As shown in the previous section, the effect of the Ca II
lines in $(U-B)_0$ and $(V-I)_0$ is significant, but their strength is
only partially responsible for the observed color variations and their
dependence on $V_{\rm neb}$ is difficult to assess. A reasonable
conclusion is that the continuum emission is also affected by viewing
angle effects.

Finally, the evidence presented here implies that both $V_{\rm neb}$
and $\Delta m^{B}_{15}$ must be taken into account when deriving
extinction values based on maximum-light colors, which is a critical
parameter in deriving distances from SNe Ia. Furthermore, as shown by
\cite{foley11}, ignoring this effect could introduce significant
biases in the determination of extinction laws ($R_V$). A larger
sample of nearby SNe Ia is urgently required to improve our calibration 
of these correlations and understand the influence of viewing angle on SN
colors.

\begin{acknowledgements}
\noindent
We would like to thank the anonymous referee for constructive comments on the text. R.C. acknowledges support by CONICYT
through Programa Nacional de Becas de Postgrado grant D-2108082 and by the Yale-Chile fellowship in astrophysics. R.C., M.H., F.F.,
and G.P. acknowledges support provided by the Millennium Center for Supernova Science through grant P10-064-F
(funded by ``Programa Bicentenario de Ciencia y Tecnolog\'ia de CONICYT'' and ``Programa Iniciativa Cient\'ifica Milenio de MIDEPLAN'').
M.H. acknowledges support provided by FONDECYT through grant 1060808. MH. and G.P. acknowledges support provided by Centro de Astrof\'\i sica FONDAP 15010003,
and Center of Excellence in Astrophysics and Associated Technologies (PFB 06). F.F. acknowledges support provided by
FONDECYT through grant 3110042. The work by K.M. is supported by World Premier International
Research Center Initiative (WPI Initiative), MEXT, Japan, and by JSPS grant in aid for Scientific
Reasearh (23740141).
G.P. acknowledges support by the Proyecto FONDECYT
11090421 and from Comit\'e Mixto ESO-Gobierno de Chile.
\end{acknowledgements}

\bibliographystyle{aa}
\bibliography{references}

\clearpage

\end{document}